\documentclass[12pt,a4paper,normalheadings,headsepline,headinclude,bibtotoc,fleqn,review]{article}
\usepackage[latin1]{inputenc}
\usepackage{amsmath}
\usepackage{booktabs}
\usepackage{array}
\usepackage{amsthm}
\usepackage{dcolumn}
\usepackage{endnotes}
\usepackage{units}
\usepackage{marvosym}
\usepackage[lines=10]{geometry}
\usepackage{longtable}
\usepackage{rotating}
\usepackage{expdlist}
\usepackage{geometry}
\usepackage{floatrow}
\usepackage{pgfplots}
\pgfmathdeclarefunction{gauss}{3}{
	\pgfmathparse{1/(#3*sqrt(2*pi))*exp(-((#1-#2)^2)/(2*#3^2))}%
}
\usetikzlibrary{decorations.pathreplacing,calligraphy,backgrounds}
\pgfplotsset{compat=1.16}

\pgfmathsetmacro\valueA{gauss(2.25, 2.25, 0.7)} 

\usepackage{wrapfig}
\usepackage{expdlist}
\usepackage{amssymb}
\usepackage{gloss}
\usepackage{nomencl}
\usepackage{natbib}

\newtheorem{prop}{Proposition}
\newtheorem{coro}{Corollary}

    \makegloss

\renewcommand{\thesection}{\arabic{section}}

\usepackage{fancyhdr} 
\begin{document}
\begin{center}\huge{Janus-Faced Technological Progress and the Arms Race in the Education of Humans and Chatbots}\footnote{I thank my former students at Zhejiang University for several conversations on the pressures in the Chinese education system.}
\end{center}

\begin{center} \textit{Wolfgang Kuhle}\\ \textit{Corvinus University of Budapest, Hungary, E-mail wkuhle@gmx.de\\ MEA, Max Planck Institute for Social Law and Social Policy, Munich, Germany}\end{center} 

\noindent\textit{\textbf{Abstract} We study the conditions under which technological advances, in combination with a lognormal wage distribution, incentivize agents into an inefficient educational arms race. Our model emphasizes that lognormal wage distributions imply that agents' wages increase exponentially in the level of their skill as well as in the level of technology. In turn, this exponential relation between skills, technology, and wages pressures agents into an exhausting race for the tails of the economy's skill distribution. Moreover, technological advances and overinvestment in education increase GDP and inequality, while welfare may decline. In an alternative interpretation, our model studies firms that invest in artificial intelligence of their chatbots and AI agents. For a wide range of specifications, firms, just like humans, have an incentive to choose corner solutions where investment is limited only by borrowing constraints.}
\vspace{0.5cm}


\section{Introduction}

In many developed countries, there is a sense that the economy, despite its GDP growth, is no longer serving the ``agents" that inhabit it. Increases in suicide rates, deaths related to substance abuse, sharp declines in fertility, widespread obesity, and the rise of populist leaders indicate that people are under significant pressure.\footnote{See \citet{Reynolds25}, \cite{case2015rising}, \cite{stiglitz2010mismeasuring}, \cite{norris2019cultural}, \cite{bongaarts2002end}, \cite{hoebel2019socioeconomic} and
\cite{furceri2012economic} for empirical evidence and a broader discussion of these points.} 

The present paper studies the conditions under which technological progress, in combination with a lognormal wage distribution, incentivizes agents into an inefficient race for the tails of the economy's skill distribution. In particular, we study the conditions under which agents' efforts to keep up with technology and one another (i) amplify inequality, (ii) incentivize agents to overinvest in education, and (iii) reduce welfare. Finally, regardless of whether technology increases or reduces welfare, we find that recent technological advances have drastically increased the opportunity cost of having children.

 We build our argument gradually. First, Section \ref{Model} shows that a lognormal wage distribution implies that earnings grow exponentially in agents' skill levels. In turn, Section \ref{Section The Pressure to Perform} shows that the exponential relationship between skills and wages induces agents to invest heavily in education. In particular, using a model without borrowing constraints, we show that technological advances can lead to a situation where the median income no longer suffices to cover agents' ex-ante optimal educational expenditures. Section \ref{Section The Pressure to Perform} thus shows that agents are ex-ante under pressure to invest heavily in education. At the same time, ex-post, agents are under pressure given that half of the population receives wages, which do not suffice to pay for their educational investment.   

Section \ref{Section Preferences/Utility Function} derives the conditions under which technological advances lower expected utility. More precisely, we show that once relative risk aversion exceeds 1, there exist technological plateaus, which provide greater utility than paths of continuous technologically driven growth.

Section \ref{Section Modern Technology, Natural Monopolies, and the Productivity Slowdown} mimics an economy where network and scale effects, in sectors such as internet search, operating systems, e-commerce or social media platforms, create large monopoly rents. In turn, agents over-invest in education in an effort to capture these monopoly rents.

In Section \ref{Calibration and Growth Accounting} we use US income data to calibrate our model. This calibration shows that, despite a doubling in per capita GDP between 1975 and 2024, agents with a coefficient of relative risk aversion exceeding 2.5 would ex-ante prefer the income distribution of 1975 over that of 2024. Moreover, to document the explosive growth in the incentive to invest in schooling, we use IQ data as a proxy for agents' skills and compute the marginal value of an additional IQ point for the income distributions of 1975 and 2024, respectively. Finally, we find that recent technological advances, together with the race for the tails of the skill distribution, have dramatically increased the opportunity costs of having children. 

Section \ref{SectionChatbot} studies the analogy between the arms race in human education and the arms race between technology firms to increase the artificial intelligence of their chatbots and AI agents. Section \ref{Section Discussion} concludes.

\paragraph{Related Literature:} The present paper aligns with the empirical literature on income and education by \cite{Aitchison1957}, \cite{Mincer1974},  \cite{Card:1999}, \cite{NealRosen2000}, \cite{Heckman2003}, \cite{Cortes2018}, who find that wage incomes are lognormally distributed. Moreover, while our calibration in Section \ref{Calibration and Growth Accounting} confirms the phenomenal returns to schooling emphasized by empirical models, the present paper emphasizes that there exist conditions where agents would be better off if the returns to education were lower. Finally, there are related empirical studies, such as \citet{Kim2026} that emphasize the arms race in education using data on a vast ``shadow education" sector providing services such as tutoring, mentoring, and other supplementary education services. 

Large parts of the theoretical literature on the formation of human capital by \cite{Becker1964}, \cite{BenPorath1967}, \cite{Card:1999}, \cite{Lochner2005}, \cite{Cunha2007}, and \cite{Bennett2015}, assume that agents face diminishing returns. This, in turn, ensures the existence of well behaved interior solutions to first-order conditions for optimal educational investment. Instead, the present paper stresses that lognormal wage distributions imply that incomes grow exponentially in agents' skill level, which can result in convex objective functions. Agents thus face a race where the next skill point is always much more valuable than the last.  

We assume that the efficient labor endowment of agents is driven by their schooling efforts. Our model thus differs from multi sector models of directed technological change by \cite{vonWeizsacker1966}, \cite{Acemoglu1998}, and \cite{Weiss2008}, where inputs, such as labor, are heterogeneous. The present paper also differs from the literature on job market signaling and overeducation by \citet{Arrow1973}, \citet{Spence1973}, \citet{Collins1979}, and \citet{Guinness2006}, who view education as a screening and signaling exercise, where agents acquire skills that are not needed to fill a certain position. In contrast, the present model assumes that increased educational efforts increase agents' productivity. 

 Finally, there is an emerging related literature, for example \citet{DECANIO2016280}, where humans can come under pressure due to the emergence of robots. In the present model, there are no robots. Instead, agents over-extend themselves in their efforts to capitalize on technological advances.

\section{Model}\label{Model} 
There is a mass one of agents. Each agent has a level of skill $y$ and a labor endowment $l(y)$. To match the empirical observation that wage income is lognormally distributed, we assume that agents' labor endowment $l(y)$ is given by: 
\begin{eqnarray}l(y)=e^{cy+b}=Ae^{cy}, \quad A:=e^{b}, \quad y\sim\mathcal{N}(\mu,\sigma), \quad c>0\label{First equation}\end{eqnarray}
Where $A$ and $c$ are technological coefficients that augment agents' skills. Finally, we normalize the wage $w=1$ so that the agents' wage earnings are:
\begin{eqnarray}
        e(y)=e^{cy+b}=Ae^{cy}\label{Wage income 1}
\end{eqnarray}
It follows from (\ref{First equation}) and (\ref{Wage income 1}) that the mean income is
\begin{eqnarray}
    E[e(y)]=Ae^{c\mu+\frac{1}{2}c^2\sigma^2} \label{Average income 1}
\end{eqnarray}
Similarly, (\ref{First equation}) and (\ref{Wage income 1}) imply that median income $\mathcal{M}$ is given by:
\begin{eqnarray}
    \mathcal{M}=Ae^{c\mu}  \label{Median income 1}
\end{eqnarray}
Combining (\ref{Average income 1}) and (\ref{Median income 1}), yields the ratio of mean to median income:
    \begin{eqnarray}
        \rho=\frac{E[e(y)]}{\mathcal{M}}=e^{\frac{1}{2}c^2\sigma^2} \implies \rho>1\label{Ratio Mean/Median}
    \end{eqnarray}
Taken together, equations (\ref{Average income 1}), (\ref{Median income 1}), and (\ref{Ratio Mean/Median}) imply:
\begin{prop}\label{Prop0}
The expected wage income of agents increases exponentially in (i) the mean skill level $\mu$ and (ii) the square of technology coefficient $c$ and (iii) in the variance/riskiness $\sigma^2$ of the skill distribution.
\end{prop}
Proposition \ref{Prop0}, together with the properties of the exponential function, yields a Corollary: 
\begin{coro} \label{Coro1}
    The expected marginal wage gain of an increase in the mean skill level $\mu$ grows exponentially in (i) the mean skill level $\mu$ and (ii) the square of technology coefficient $c$ and (iii) in the variance/riskiness $\sigma^2$ of the skill distribution.
\end{coro}

In order to bring out the implications of Proposition \ref{Prop0} and Corollary \ref{Coro1}, Sections \ref{Section The Pressure to Perform}-\ref{Section Modern Technology, Natural Monopolies, and the Productivity Slowdown} now develop three different model variants. These variants emphasize how exponential returns to skills can yield unfavorable results for the inhabitants of the economy.

\section{Pressure to Perform I: Median Income Falling Below Educational Outlays}\label{Section The Pressure to Perform} 

In this section, we examine agents' incentive to invest in their skills. That is, agents can invest an amount $I$ in schooling to improve their expected skill $\mu(I)$, respectively, to maximize their expected earnings net of educational investment. For simplicity, we abstract from borrowing constraints. Recalling (\ref{First equation}) and (\ref{Wage income 1}) we have:
\begin{eqnarray}
    \max_{I}\{E[e(y)]-I\}=\max_I\{Ae^{c\mu(I)+\frac{1}{2}c^2\sigma^2}-I\} \label{Max Education}
\end{eqnarray}
where (\ref{Max Education}) yields a FOC:
\begin{eqnarray}
Ae^{c\mu+\frac{1}{2}c^2\sigma^2}c\frac{\partial\mu}{\partial I}-1=0 \label{FOC Pressure to perform}
\end{eqnarray}
It is useful to note two properties of (\ref{FOC Pressure to perform}). First, increases in mean skill $\mu$ exponentially increase marginal earnings. That is, there tend to exist corner solutions, with run-away investment in education, for many functional forms $\mu(I)$. Second, to further interpret (\ref{FOC Pressure to perform}), it is useful to consider the special case $\mu=ln(I)$, where (\ref{FOC Pressure to perform}) has an interior optimum solution for as long as $c<1$,\footnote{Our calibration in Section \ref{Calibration and Growth Accounting} suggests that indeed $c<1$.} which writes: \begin{eqnarray}
    Ae^{c\mu+\frac{1}{2}c^2\sigma^2}c=I \quad \Leftrightarrow \quad Ae^{\frac{1}{2}c^2\sigma^2}I^{(c-1)}c=1\quad \Leftrightarrow \quad
    I=(Ae^{\frac{1}{2}c^2\sigma^2}c)^{^{\frac{1}{(c-1)}}}\label{FOC Pressure to perform 2}
\end{eqnarray}
Using (\ref{Median income 1}), we can rewrite (\ref{FOC Pressure to perform 2}), and compare ex-ante optimal educational investment with median income:
\begin{eqnarray}
    \mathcal{M}-I=I(\frac{1}{c}e^{-\frac{1}{2}c^2\sigma^2}-1) \label{Median Net Income}
\end{eqnarray}
Where (\ref{Median Net Income}) implies:
\begin{prop}\label{Prop1}
In the limit where $c\rightarrow1$ median income $\mathcal{M}$ falls short of the ex-ante optimal level of educational investment $I$. 
\end{prop}
Proposition \ref{Prop1} emphasizes how increases in the technological level $c$ ex-ante incentivize agents' to invest heavily in education. Ex-post, however, this high educational investment only pays off for a small fraction of all agents, while ex-post the majority receives incomes that do not suffice to cover the ex-ant optimal educational outlays. Thus, Proposition \ref{Prop1} shows that ex-ante agents are under pressure to invest heavily in education. At the same time, over half of all agents find that their ex-post wages do not cover their educational investment. That is, more than half of all agents are ex-post worse off than they would have been had they not invested in education.

\section{Pressure to Perform II: Preferences over Technology}\label{Section Preferences/Utility Function}
This section examines agents' preferences regarding the level of technology $c$, to offer another perspective on how technological progress my pressure agents. Using CRRA preferences over consumption $C$, recalling the distribution of labor (\ref{First equation}) and wage income (\ref{Wage income 1}), to compute agents' expected utility:
\begin{eqnarray}
    E[U]=E[\frac{1}{1-\phi}C^{1-\phi}],\quad C=e(y)=e^{cy+b} \label{Utility}
\end{eqnarray}
Substituting agents' wage income (\ref{Wage income 1}) into (\ref{Utility}) yields:
\begin{eqnarray}
    E[U]=\frac{1}{1-\phi}e^{(1-\phi)(c\mu+b)+\frac{1}{2}(1-\phi)^2c^2\sigma^2}
\end{eqnarray}
such that technology $c$ maximizes expected utility if:
\begin{eqnarray}
    \frac{\partial E[U]}{\partial c}=e^{(1-\phi)(c\mu+b)+\frac{1}{2}(1-\phi)^2c^2\sigma^2}[\mu+(1-\phi)c\sigma^2]=0 \label{FOC OPT Technology}
\end{eqnarray}
Where (\ref{FOC OPT Technology}) implies that optimal technology is a corner solution for as long as $0<\phi<1$. That is, as long as the coefficient of relative risk aversion is less than one, run-away technological growth is optimal. 

For risk aversion $\phi>1$, there exists an interior optimum level, respectively, an optimal plateau for technology:\footnote{The sufficient condition for an optimum $\frac{\partial^2 E[U]}{\partial c^2}_{|c=c^*}=e^{(1-\phi)(c\mu+b)+\frac{1}{2}(1-\phi)^2c^2\sigma^2}(1-\phi)\sigma^2<0$ is satisfied as long as $\phi>1$.}
\begin{eqnarray}
    c^*=\frac{\mu}{(\phi-1)\sigma^2},   \quad \phi>1
\end{eqnarray}
Taken together, we thus have:
\begin{prop}\label{prop2}
    Technological advances always increase expected utility if $\phi<1$. If $\phi>1$ there exists an ex-ante optimal technological plateau $c^*=\frac{\mu}{(\phi-1)\sigma^2}$.
\end{prop}
That is, while mean incomes and thus GDP are monotonously increasing in the technological level $c$, Proposition \ref{prop2} shows that for coefficients of relative risk aversion exceeding 1,\footnote{See \cite{FriendBlume1975}, \cite{Pindyck1988},
\cite{FrenchSchwertStambaugh1987},
\cite{MehraPrescott1985},
\cite{KandelStambaugh1991},
\cite{Kaplow2005},
\cite{Chetty2006}, for references arguing that coefficients of relative risk aversion exceed 1.} agents' ex-ante utility eventually starts to fall once $c$ exceeds the optimal level $c^*$. 

\section{Pressure to Perform III: The Bell Curve Competition}\label{Section Modern Technology, Natural Monopolies, and the Productivity Slowdown}
Increasing returns to scale and network effects enable modern technology firms to capture large monopoly rents in industries such as operating systems, search engines, social media platforms, or software engineering tools. These monopoly rents imply that parts of the income of entrepreneurs and their employees are derived from the fact that they operate a monopoly. 

In the following, the coefficient $\alpha\in[0,1]$ measures the degree to which workers' incomes depend on their actual productivity. In turn, $1-\alpha$ captures the share of income that is merely due to the fact that one worker manages to out-compete others and capture monopoly rents.

Taking the perspective of a particular agent $i$, expected wages net of educational investment $I_i$ are: 
\begin{eqnarray}
    &&E[e(y_i)-I_i]=E[Ae^{\alpha c y_i+(1-\alpha)c(y_i-\mu)}]-I_i, \label{Productivity slowdown Base Equation 1}\\&& \alpha\in[0,1], \quad y_i\sim\mathcal{N}(\mu_{i},\sigma), \quad \mu_{i}=\mu_{i}(I_i), \quad \mu=\int_{[0,1]}\mu_{i}di \nonumber
\end{eqnarray}
Here, the term $\alpha c y_i$ in (\ref{Productivity slowdown Base Equation 1}) represents the share of agents' productivity that depends on their individual skill. In turn, the term $(1-\alpha)c(y_i-\mu)$ represents the share of their productivity that stems from the fact that they are ``quicker on their feet" than the rest of the population. Put differently, $(1-\alpha)c(y_i-\mu)$ reflects that agents with above average skills $y_i$ are more successful in capturing the economy's monopoly rents. 
Equation (\ref{Productivity slowdown Base Equation 1}) rewrites
\begin{eqnarray}
   E[e(y_i)-I_i]=Ae^{c\mu_{i}-(1-\alpha)c\mu+\frac{1}{2}c^2\sigma^2}-I_i
\end{eqnarray}
In turn, the FOC for private investment in education reads: 
\begin{eqnarray}
    \frac{\partial (E[e(y_i)]-I_i)}{\partial I_i}= Ae^{c\mu_{i}-(1-\alpha)c\mu+\frac{1}{2}c^2\sigma^2}c\frac{\partial \mu_i}{\partial I_i}-1=0
\label{FOC Private}\end{eqnarray}
The condition (\ref{FOC Private}) for optimal individual investment $I_i$ reflects that agents take the mean skill $\mu$ of the population as given. 

To show that agents over-invest in education, we compute the socially optimal level of investment. Taking the social planner's view, we note that in equilibrium we have $\mu_{i}=\mu$ and $I_i=I$, such that the FOC for the socially optimal level of educational investment is:
\begin{eqnarray}
     \frac{\partial (E[e(y)]-I)}{\partial I}=Ae^{c\mu-(1-\alpha)c\mu+\frac{1}{2}c^2\sigma^2}\alpha c\frac{\partial \mu}{\partial I}-1=0\label{FOC SOP}
\end{eqnarray}
Comparing agents' private investment choice (\ref{FOC Private}) with socially optimal investment (\ref{FOC SOP}), indicates that: 
\begin{prop} \label{prop3}
    Agents over-invest in education for as long as $\alpha\in(0,1)$.
\end{prop}
In one interpretation, Proposition \ref{prop3} shows that economies, where many technology firms operate as monopolies, i.e., where $1-\alpha$ is high, induce an inefficient competition for education. Moreover, for any given level of educational investment, decreases in $\alpha$, respectively, increases in $1-\alpha$, induce a productivity slowdown.\footnote{That is, mean labor productivity is given by $Ae^{c\mu-(1-\alpha)c\mu+\frac{1}{2}c^2\sigma^2}$. Thus, productivity decreases once $\alpha$ does, respectively, productivity falls once $1-\alpha$ increases.} That is, as the competition between agents to capture monopoly rents in the tech industry increases, workers, and thus the economy become less productive.

\section{Calibration}\label{Calibration and Growth Accounting}
To identify the coefficients of our baseline model (\ref{First equation})-(\ref{Median income 1}), we solve (\ref{Average income 1}) and (\ref{Median income 1}) for $A$ and $c$:
\begin{eqnarray}
  A=\mathcal{M}e^{-c\mu}, \quad  c^2=\frac{ln(\rho)}{\frac{1}{2}\sigma^2}, c>0, \implies c=\sqrt{\frac{ln(\rho)}{\frac{1}{2}\sigma^2}}.\label{Cali1}
\end{eqnarray}
\begin{eqnarray}
        \rho=\frac{E[e(y)]}{\mathcal{M}}=e^{\frac{1}{2}c^2\sigma^2}, \quad A=e^b.\label{Ratio Mean/Medianrep}
    \end{eqnarray}
To compute $A$ and $c$, we use 2024 US census data (Table H-11), for mean and median household income. 
Moreover, we use IQ as a proxy for the skill of agents $y$. Most empirical studies suggest that IQ is normally distributed such that $y\sim\mathcal{N}(100,15)$. Substituting these values into (\ref{First equation})-(\ref{Median income 1}), respectively (\ref{Cali1}) and (\ref{Ratio Mean/Medianrep}) yields: 
\begin{enumerate}
    \item 1975: $\mathcal{M}=58000$, $E[e]=68000$, $c_{1975}=0.0376$, $b_{1975}=7.2$, $A_{1975}=1339.4$
    \item 2024: $\mathcal{M}=83000$, $E[e]=121000$, $c_{2024}=0.0579$, $b_{2024}=5.5$, $A_{2024}=244.7$
\end{enumerate}
Comparison of coefficients for years $1975$ and $2024$ indicates that the technology coefficient $c$, which directly augments agents' skills $y$, has grown by more than $50$\%. At the same time, agents ``baseline productivity" $A$ has fallen by more than $80$\%. That is, while mean and median income have grown, agents' incomes in 2024 are more risky than they were in 1975. Put differently, earnings have disproportionally moved into the upper tail of the skill distribution.

\paragraph{Preferences over the Income Distribution:} To emphasize the increased risk in agents' earnings, we use our model from Section \ref{Section Preferences/Utility Function} in Appendix \ref{Appendix1}, and show that agents with relative risk aversion of $\phi^*=2.5$ are indifferent between the income distribution of 1975 and 2024. Put differently, despite a doubling in mean incomes, expected utility for agents with risk aversions exceeding $2.5$ is lower in 2024 than it was in 1975.

\paragraph{Expected Value of one Additional IQ Point:}
Recalling (\ref{Average income 1}) expected wage earnings are $E[e(y)]=Ae^{c\mu+\frac{1}{2}c^2\sigma^2}$. Hence, an increase in expected IQ/skill from $\mu$ to $\mu+1$ increases expected earnings by $\Delta:=E_{\mu+1}[e(y)]-E_{\mu+1}[e(y)]=Ae^{c\mu+\frac{1}{2}c^2\sigma^2}[e^{c}-1]$. 

\begin{center}
\begin{tikzpicture}
\begin{axis}[
    xlabel={$IQ/Skill$},
    ylabel={$\ln(\Delta)$},
    xmin=0, xmax=170,
    ymin=0, ymax=1e6,  
    legend pos=north west,
    grid=major,
    width=12cm,
    height=8cm,
    ymode=log,  
]

\addplot[blue, thick, domain=0:170, samples=100] 
    {1339 * exp(0.0376 * x + 0.5 * (0.0376)^2 * 15^2) * (exp(0.0376) - 1)};
\addlegendentry{$1975$}

\addplot[red, thick, domain=0:170, samples=100] 
    {244.7 * exp(0.0579 * x + 0.5 * (0.0579)^2 * 15^2) * (exp(0.0579) - 1)};
\addlegendentry{2024}
\end{axis}
\end{tikzpicture}
\end{center}
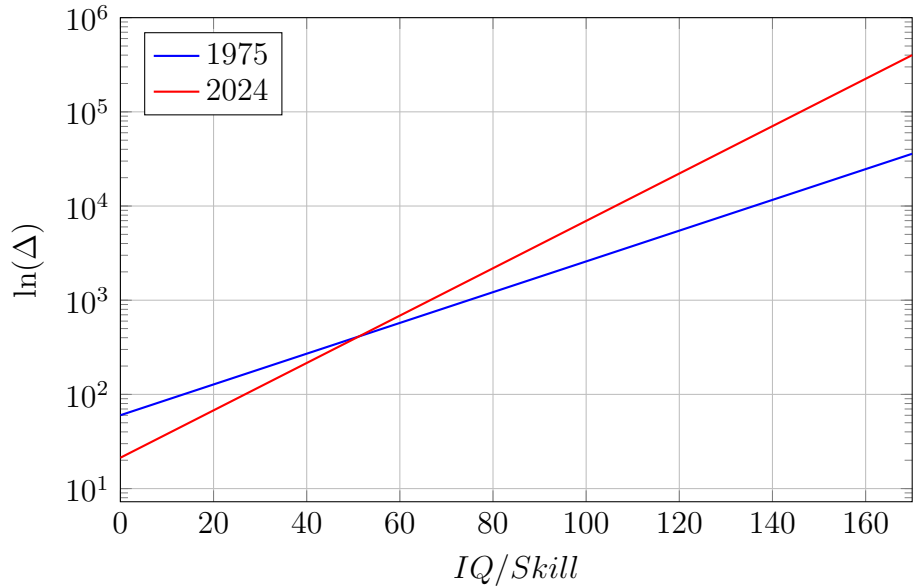
\textbf{Figure 1:}\textit{ The blue and the red curve represent the logarithm of the expected wage increase (in 2024 USD), which is generated by one additional IQ point, for the years 1975 and 2024 respectively.}

Diagram 1 shows that the ex-ante pressure to invest in education has gone up drastically for all agents with an IQ/skill larger than 50. To grasp the magnitude of this increase, Table 1 (below) computes the value of an additional IQ/skill point for different levels of IQ. For example, row 115 of Table 1 shows that the ex-ante incentive to invest in education has almost quadrupled for agents with an IQ of 115. Assuming a 40 year work-life, the inflation adjusted marginal value for one IQ point has gone from 181600\$ to 663040\$ for agents with a skill level of 115. 

\begin{table}[h]
\centering
\begin{tabular}{|c|c|c|c|}
\hline
IQ/Skill & 1975 & 2024 & 2073 \\
\hline
70 & 836.17 & 1224.43 & 5202.92 \\
85 & 1469.73 & 2918.23 & 19800.74 \\
100 & 2583.33 & 6955.10 & 75355.67 \\
115 & 4540.69 & 16576.33 & 286781.09 \\
130 & 7981.12 & 39506.91 & 1091402.85 \\
145 & 14028.33 & 94158.13 & 4153552.06 \\
160 & 24657.44 & 224410.19 & 15807173.97 \\
175 & 43340.12 & 534844.26 & 60157365.43 \\
\hline
\end{tabular}
\end{table}
\textbf{Table 1:} \textit{Increase in annual salaries for one additional IQ/skill point in 2024 Dollars. The column for the year 2073 extrapolates the percentage changes in coefficients $A$ and $c$, which occurred between 1975 and 2024.}

Together, the rows of Table 1 show that technology has drastically amplified agents' incentive to invest in education between the years 1975 and 2024. Finally, Column 2073 extrapolates the changes between the years 1975 and 2024 to the year 2073. Column 2073 may thus be viewed as the earnings gains that agents, who decide on their education today, can expect if current trends persist.

\paragraph{Interpretation:} Graph 1 and Table 1 allow for several interpretations. Both, the decrease in the ``baseline productivity" $A$ as well as the increase in the coefficient $c$, which augments agents' skill $y$, may be viewed as the expression of an increasingly meritocratic system where, on average, large efforts are rewarded with large earnings. Graph 1 shows this: the intercept of the earnings curves moved lower between 1975 and 2024, while the slope of logarithmic curve increased. 

In an alternative interpretation, Table 1 shows that the value of an additional skill point increases exponentially in the level of agents' skills. In turn, as  discussed in Sections \ref{Section The Pressure to Perform} and \ref{Section Modern Technology, Natural Monopolies, and the Productivity Slowdown}, this exponential relation between skill and wage can incentivize over-investment in education. That is, agents find themselves in an exhausting race without finish-line, where their earnings are mainly driven by the last few skill points that they acquire.

In a third interpretation, Table 1 represents the opportunity cost of having family and children. College education is typically assumed to add between 1 and 5 IQ points per year of enrollment. Hence, if young men and women with children forego 2 or 3 years of college, assuming a 40 year work-life, columns 2024 and 2073 in Table 1 quickly predict millions in foregone life-cycle earnings for anyone exceeding the mean skill of 100. Taking this view, Table 1 suggests that the financial incentives for child bearing offered in many western countries are orders of magnitude smaller than the current, technologically driven, increases in the opportunity cost of having children.

\section{Training of Chatbots and AI Agents}\label{SectionChatbot}

Next to investment in human intelligence, firms are currently investing heavily to increase the artificial intellignece of their chatbots and AI agents. If one accepts the analogy between the training of humans and AI agents, then the planned AI capex of roughly 700 billion USD for the year 2026 by Amazon, Google, Microsoft, Meta and Oracle, illustrate the extreme pressure in the ``education sector." Put differently, firms like Google with a 2025 cashflow of 165 billion USD are under pressure to issue debt to fund a planned AI capex of 185 billion USD for the year 2026. That is, firms, just like humans, go to the limits of their financial means in an effort to develop and train/educate AI models which are in the extreme tails of the AI skill distribution.    
  
Following this interpretation further, Section \ref{Section The Pressure to Perform} presents a model in which firms, rather than agents, choose how much to invest in the education of their AI agent. In turn, if the returns to chatbot skill/IQ follow a log-normal distribution we once again face a situation where returns increase exponentially in the AI model's skill. Hence, for a wide range of model specifications, as we have discussed in Section  \ref{Section The Pressure to Perform}, firms choose corner solutions, where they invest as much as they can in the education of their AI model. In turn, ex-post, for most firms these investments do not pay off, resulting in a situation akin to the student loan crises in the market for AI related loans. 

Similarly, Section \ref{Section Modern Technology, Natural Monopolies, and the Productivity Slowdown} may be viewed as a model where scale and network effects amplify the competitive pressure such that educational investment is ex-ante inefficiently high. That is, if chatbots compete for traffic, then one platform, which is on average just a little better than the others, will capture a disproportionate share of the market. In turn, firms have an ex-ante incentive to overinvest in their AI models just to stay ahead of the competition. 
 
\section{Discussion}\label{Section Discussion}

This paper emphasizes that lognormal income distributions imply that agents' wages increase exponentially in their skill. These exponential returns, in turn, create an unbalanced situation in which agents, by several metrics, overinvest in education in an effort to reach the extreme tails of the economy's skill distribution.

Technological progress is Janus-faced in the present model. On the one hand, technology increases productivity and GDP. On the other hand, these gains in productivity are concentrated in the upper tail of the skill distribution. Technology thus amplifies the educational arms race among agents who over-extend themselves, trying to keep up with technology and one another.     

To capture the pressure that the economy places on agents, we considered three arguments. First, we show that technological advances can incentivize agents to choose a level of educational investment which exceeds ex-post median income. That is, agents are ex-ante under pressure to invest heavily in education. At the same time, upon graduation, more than half of all agents are under pressure because their wages do not cover ex-ante optimal educational outlays. 

Second, we show that agents with a coefficient of relative risk aversion larger than one prefer technological plateaus, i.e steady states where the technology that augments skills does not advance further, over other paths where technology continuously improves mean incomes and GDP. In order to study how ``far" the current technology is from such an optimal plateau, we calibrate our model in Section \ref{Calibration and Growth Accounting}. This calibration indicates that, despite a doubling in per capita GDP between 1975 and 2024, agents with a coefficient of relative risk aversion higher than 2.5 would ex-ante prefer the income distribution of 1975 over that of 2024. 

Third, we considered a model where technological advances create economies of scale that allow companies to capture monopoly rents. These rents, in turn, amplify the educational among agents, who now compete for jobs in the ``tech sector."   

Fourth, we discussed that the arms race in human education is similar to the arms race between technology firms, such as Amazon, Google, Microsoft and Meta, who invest heavily in the development and training of chatbots and AI agents. That is, despite having some of the largest cash flows in corporate history, competition is such that these firms have to borrow to fund the education of their AI models. That is, investment in the education of humans and AI agents both exhibit patterns of behavior that are consistent with corner solutions.   

Finally, we find that technological change together with the arms race in education have drastically increased the opportunity cost of having family and children. That is, the child benefits introduced in many Western countries are dramatically smaller than recent technologically driven increases in the opportunity cost of having children.

\newpage
\addcontentsline{toc}{section}{References}
\markboth{References}{References}
\bibliographystyle{apalike}
\bibliography{References}

\begin{thebibliography}{}

\bibitem[Acemoglu, 1998]{Acemoglu1998}
Acemoglu, D. (1998).
\newblock Why do new technologies complement skills? directed technical change and wage inequality.
\newblock {\em The Quarterly Journal of Economics}, 113(4):1055--1089.

\bibitem[Aitchison and Brown, 1957]{Aitchison1957}
Aitchison, J. and Brown, J. A.~C. (1957).
\newblock {\em The Lognormal Distribution, with Special Reference to Its Uses in Economics}.
\newblock Cambridge University Press, Cambridge.

\bibitem[Arrow, 1973]{Arrow1973}
Arrow, K.~J. (1973).
\newblock Higher education as a filter.
\newblock {\em Journal of Public Economics}, 2(3):193--216.

\bibitem[Becker, 1964]{Becker1964}
Becker, G.~S. (1964).
\newblock {\em Human Capital: A Theoretical and Empirical Analysis, with Special Reference to Education}.
\newblock University of Chicago Press, Chicago, 1 edition.

\bibitem[Ben-Porath, 1967]{BenPorath1967}
Ben-Porath, Y. (1967).
\newblock The production of human capital and the life cycle of earnings.
\newblock {\em Journal of Political Economy}, 75(5):352--365.

\bibitem[Bennet et~al., 2015]{Bennett2015}
Bennet, D., Ngum, N., and O'Connell, B. (2015).
\newblock Human capital, learning, and the accumulation of abilities.
\newblock {\em Journal of Economic Theory}, 160:1--29.

\bibitem[Bongaarts, 2002]{bongaarts2002end}
Bongaarts, J. (2002).
\newblock The end of the fertility transition in the developed world.
\newblock {\em Population and Development Review}, 28(3):419--443.

\bibitem[Card, 1999]{Card:1999}
Card, D. (1999).
\newblock The causal effect of education on earnings.
\newblock In Ashenfelter, O. and Card, D., editors, {\em Handbook of Labor Economics}, volume~3, chapter~30, pages 1801--1863. Elsevier, 1 edition.

\bibitem[Case and Deaton, 2015]{case2015rising}
Case, A. and Deaton, A. (2015).
\newblock Rising morbidity and mortality in midlife among white non-hispanic americans in the 21st century.
\newblock {\em Proceedings of the National Academy of Sciences}, 112(49):15078--15083.

\bibitem[Chetty, 2006]{Chetty2006}
Chetty, R. (2006).
\newblock A new method of estimating risk aversion.
\newblock {\em American Economic Review}, 96(5):1821--1840.

\bibitem[Collins, 1979]{Collins1979}
Collins, R. (1979).
\newblock {\em The Credential Society}.
\newblock Academic Press.

\bibitem[Cortes et~al., 2018]{Cortes2018}
Cortes, G.~M., Lochner, L., Park, Y., and Shin, Y. (2018).
\newblock Wage dynamics and returns to unobserved skills.
\newblock NBER Working Paper Series 24220, National Bureau of Economic Research.

\bibitem[Cunha and Heckman, 2007]{Cunha2007}
Cunha, F. and Heckman, J.~J. (2007).
\newblock The technology of skill formation.
\newblock {\em American Economic Review}, 97(2):31--47.

\bibitem[DeCanio, 2016]{DECANIO2016280}
DeCanio, S.~J. (2016).
\newblock Robots and humans complements or substitutes?
\newblock {\em Journal of Macroeconomics}, 49:280--291.

\bibitem[French et~al., 1987]{FrenchSchwertStambaugh1987}
French, K.~R., Schwert, G.~W., and Stambaugh, R.~F. (1987).
\newblock Expected stock returns and volatility.
\newblock {\em Journal of Financial Economics}, 19(1):3--29.

\bibitem[Friend and Blume, 1975]{FriendBlume1975}
Friend, I. and Blume, M.~E. (1975).
\newblock The demand for risky assets.
\newblock {\em The American Economic Review}, 65(5):900--922.

\bibitem[Furceri et~al., 2012]{furceri2012economic}
Furceri, D., Loungani, P., and Ostry, J.~D. (2012).
\newblock Economic growth and the rise of political extremism: Theory and evidence.
\newblock IMF Working Papers 12/220, International Monetary Fund.

\bibitem[Heckman et~al., 2003]{Heckman2003}
Heckman, J.~J., Lochner, L.~J., and Todd, P.~E. (2003).
\newblock Fifty years of mincer earnings regressions.
\newblock NBER Working Paper Series 9732, National Bureau of Economic Research.

\bibitem[Hoebel et~al., 2019]{hoebel2019socioeconomic}
Hoebel, J., Kuntz, B., Hegerl, U., Stein, J., Rüter, A., Mayer, S., Hauschildt, J., Fuchs, J., Helmert, U., Jacobi, C., et~al. (2019).
\newblock Socioeconomic inequalities in the rise of adult obesity: Trend analyses of population-based national health survey data in germany, 1999-2012.
\newblock {\em Frontiers in Public Health}, 7:146.

\bibitem[Kandel and Stambaugh, 1991]{KandelStambaugh1991}
Kandel, S. and Stambaugh, R.~F. (1991).
\newblock Asset returns and intertemporal preferences.
\newblock {\em Journal of Monetary Economics}, 27(1):39--71.

\bibitem[Kaplow, 2005]{Kaplow2005}
Kaplow, L. (2005).
\newblock The value of a statistical life and the coefficient of relative risk aversion.
\newblock {\em Journal of Risk and Uncertainty}, 31(1):23--34.

\bibitem[Kim, 2026]{Kim2026}
Kim, T. (2026).
\newblock The impact of shadow education expenditures on fertility rates in {South Korea}.
\newblock {\em Journal of Population Economics}, 39(1):5.
\newblock Published online: 04 February 2026.

\bibitem[Lochner and Monge-Naranjo, 2005]{Lochner2005}
Lochner, L. and Monge-Naranjo, A. (2005).
\newblock The nature of credit constraints and human capital.
\newblock {\em American Economic Review}, 95(5):1422--1454.

\bibitem[McGuinness, 2006]{Guinness2006}
McGuinness, S. (2006).
\newblock Overeducation in the labour market.
\newblock {\em Journal of Economic Surveys}, 20(3):387--418.

\bibitem[Mehra and Prescott, 1985]{MehraPrescott1985}
Mehra, R. and Prescott, E.~C. (1985).
\newblock The equity premium: A puzzle.
\newblock {\em Journal of Monetary Economics}, 15(2):145--161.

\bibitem[Mincer, 1974]{Mincer1974}
Mincer, J. (1974).
\newblock {\em Schooling, Experience, and Earnings}.
\newblock National Bureau of Economic Research, New York.

\bibitem[Neal and Rosen, 2000]{NealRosen2000}
Neal, D. and Rosen, S. (2000).
\newblock Theories of the distribution of earnings.
\newblock In Atkinson, A.~B. and Bourguignon, F., editors, {\em Handbook of Income Distribution}, volume~1, chapter~7, pages 379--427. Elsevier, 1 edition.

\bibitem[Norris and Inglehart, 2019]{norris2019cultural}
Norris, P. and Inglehart, R. (2019).
\newblock {\em Cultural Backlash: Trump, Brexit, and Authoritarian Populism}.
\newblock Cambridge University Press.

\bibitem[Pindyck, 1988]{Pindyck1988}
Pindyck, R.~S. (1988).
\newblock Risk aversion and determinants of stock market behavior.
\newblock {\em The Review of Economics and Statistics}, 70(2):183--190.

\bibitem[Reynolds, 2025]{Reynolds25}
Reynolds, N. (2025).
\newblock The broad decline in health and human capital of americans born after 1947.
\newblock {\em American Economic Review: Insights}, 7(2):141--59.

\bibitem[Spence, 1973]{Spence1973}
Spence, M. (1973).
\newblock Job market signaling.
\newblock {\em Quarterly Journal of Economics}, 87(3):355--374.

\bibitem[Stiglitz et~al., 2010]{stiglitz2010mismeasuring}
Stiglitz, J.~E., Sen, A., and Fitoussi, J.-P. (2010).
\newblock {\em Mismeasuring Our Lives: Why GDP Doesn't Add Up}.
\newblock The New Press, New York.

\bibitem[Weiss, 2008]{Weiss2008}
Weiss, M. (2008).
\newblock Skill-biased technological change: Is there hope for the unskilled?
\newblock {\em Economics Letters}, 98(3):439--445.

\bibitem[Weizsaecker, 1966]{vonWeizsacker1966}
Weizsaecker, C. C.~v. (1966).
\newblock Tentative notes on a two-sector model with induced technical progress.
\newblock {\em The Review of Economic Studies}, 33(3):245--251.

\end{thebibliography}

\newpage

\appendix
\section{Relative Utility}\label{Appendix1}
Using our model from Section \ref{Section Preferences/Utility Function} we can compare the ratio of ex-ante expected utilities for agents born into the income distributions of the years 2024 and 1975 respectively:
\begin{eqnarray}
    \frac{E[U_{2024}]}{E[U_{1975}]}=e^{(1-\phi)(c_{2024}-c_{1975})\mu+b_{2024}-b_{1975}+\frac{1}{2}(1-\phi)(c^2_{2024}-c^2_{1975})\sigma^2}\geqq 1
\end{eqnarray}
Hence, for $\phi>1$, the technology induced income distribution for the year 1975 is yielding higher expected utility than the income distribution of the year 2024 if:  
\begin{eqnarray}
    ln(\frac{E[U_{2024}]}{E[U_{1975}]})=(1-\phi)\Big((c_{2024}-c_{1975})\mu+b_{2024}-b_{1975}+\frac{1}{2}(1-\phi)(c^2_{2024}-c^2_{1975})\sigma^2\Big)>0\nonumber\\
    \Leftrightarrow    (c_{2024}-c_{1975})\mu+b_{2024}-b_{1975}+\frac{1}{2}(1-\phi)(c^2_{2024}-c^2_{1975})\sigma^2<0\nonumber
\end{eqnarray}
The cutoff risk aversion is thus:
\begin{eqnarray}
    \phi^*=\frac{(c_{2024}-c_{1975})\mu+b_{2024}-b_{1975}+\frac{1}{2}(c^2_{2024}-c^2_{1975})\sigma^2}{\frac{1}{2}(c^2_{2024}-c^2_{1975})\sigma^2}
\end{eqnarray}

\end{document}